\documentclass[aps, prl,amsmath,amssymb, reprint, superscriptaddress]{revtex4-2}

\usepackage{xcolor}
\usepackage{graphicx}
\usepackage{dcolumn}
\usepackage{bm}

\newcommand{\pd}{\partial}

\begin{document}

\preprint{APS/123-QED}

\title{Emission of fast-propagating spin waves by \\an antiferromagnetic domain wall driven by spin current}

\author{Roman V. Ovcharov}
\affiliation{ 
Department of Physics, University of Gothenburg, Gothenburg 41296, Sweden
}

\author{B.A. Ivanov}
\affiliation{
Institute of Magnetism of NASU and MESU, Kyiv 03142, Ukraine
}
\affiliation{
William H. Miller III Department of Physics and Astronomy, Johns Hopkins University, Baltimore, Maryland 21218, USA
}

\author{Johan \AA kerman}
\affiliation{ 
Department of Physics, University of Gothenburg, Gothenburg 41296, Sweden
}
\affiliation{
Center for Science and Innovation in Spintronics, Tohoku University, Sendai 980-8577, Japan
}
\affiliation{
Research Institute of Electrical Communication, Tohoku University, Sendai 980-8577, Japan
}

\author{Roman S.  Khymyn}%
\affiliation{ 
Department of Physics, University of Gothenburg, Gothenburg 41296, Sweden
}

\date{\today}

\begin{abstract}
Antiferromagnets (AFMs) have great benefits for spintronic applications such as high frequencies (up to THz), high speeds (up to tens of km/s) of magnetic excitations, and field-free operation. Advanced devices will require high-speed propagating spin waves (SWs) as signal carriers, i.e., SWs with high k-vectors, the excitation of which remains challenging. We show that a domain wall (DW) in anisotropic AFM driven by the spin current can be a source of such propagating SWs with high frequencies and group velocities. In the proposed generator, the spin current, with polarization directed along the easy anisotropy axis, excites the precession of the N\'eel vector within the DW. The threshold current is defined by the value of the anisotropy in the hard plane, and the frequency of the DW precession is tuneable by the strength of the spin current. We show that the above precession of spins inside the DW leads to robust emission of high-frequency propagating SWs into the AFM strip with very short wavelengths comparable to the exchange length, which is hard to achieve by any other method.
\end{abstract}

\maketitle

\textit{Introduction.---}Spin-transfer-torque and spin-Hall auto-oscillators (AOs) based on ferromagnetic materials (FMs) are well-established devices in modern spintronics and have a great potential for advanced signal and data processing~\cite{demidov2012magnetic, demidov2014nanoconstriction, slavin2009nonlinear, chen2016spin}. For example, thanks to their highly nonlinear behavior, they are promising in neuromorphic computing applications, such as image or sound recognition~\cite{grollier2020neuromorphic, romera2018vowel, zahedinejad2020two}. Such complex tasks require large arrays of strongly mutually coupled AOs that can be achieved by direct exchange, magneto-dipolar interactions, or spin waves (SWs) propagating between individual AOs~\cite{kaka2005mutual, mancoff2005phase, sani2013mutually, locatelli2015efficient, houshang2016spin, lebrun2017mutual}. The latter has special advantages since SWs can carry signals on large distances and be additionally processed in the inter-AOs space~\cite{awad2017long, zahedinejad2020two, kumar2023robust}. 
Despite the above benefits, the FM AOs have significant drawbacks, such as the necessity of externally applied strong magnetic field, low operational frequencies, which are usually limited by a few tens of GHz~\cite{bonetti2009spin}, and low velocity of the emitted SWs, which are of the order of 1 km/s.

\begin{figure}[hbt!]
\centering
	\includegraphics[width=\linewidth]{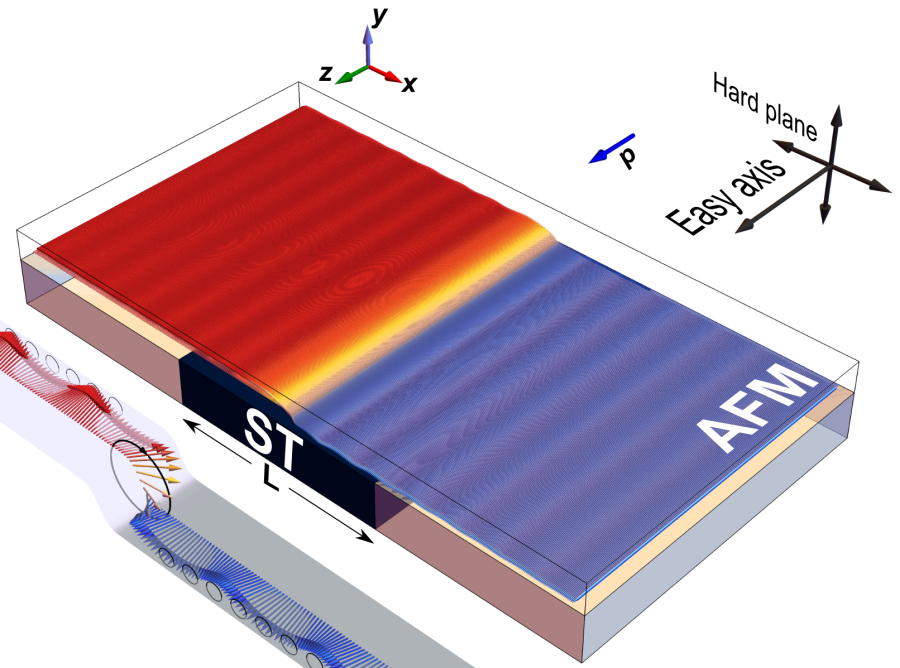}
	\caption{Schematic representation of the proposed ultrashort SW generator. The spin torque source with a width $L$ is positioned at the device's center beneath the AFM DW. The spin current polarization $\mathbf{p}$ aligns with the easy axis. The spin current application induces a precession of the N\'eel vector within the DW in the hard plane with n-fold symmetry. This precession results in the emission of SWs, as schematically shown in the lower-left corner.}
	\label{fig:schema}
\end{figure}

Recently, it was proposed to use antiferromagnetic materials (AFM) instead of FM to eliminate the above issues~\cite{cheng2016terahertz, khymyn2017antiferromagnetic, sulymenko2017terahertz, johansen2017spin, gomonay2018antiferromagnetic, puliafito2019micromagnetic, lee2019antiferromagnetic, troncoso2019antiferromagnetic, parthasarathy2021precessional}. AFM AOs can operate in the THz frequency range and do not require an external magnetic field due to the well-known feature of the AFM spin dynamics – the utilization of the internal exchange field or so-called – exchange amplification~\cite{gomonay2014spintronics, baltz2018antiferromagnetic, galkina2018dynamic}. The velocities of the SWs in AFMs can reach dozens of km/s~\cite{hortensius2021coherent}, which is promising for the fast signal/data transduction between AOs. However, substantially short wavelengths of the excited magnons are required to achieve such high velocities. The dispersion relation for the propagating SWs in AFM reads as $\omega=\sqrt{\omega_0^2 + c^2 k^2}$ (where $\omega_0$ is the frequency of AFM resonance (AFMR), $c$ – maximum group velocity of magnons and $k$ denotes a wavevector). Thus, the group SW velocity, $v_{gr}=\pd \omega/\pd k$, tends to zero for a small $k$, and one is interested in the case $k \gtrsim \omega_0/c$, which corresponds to the wavelength of a few tens of nanometers for the typical AFMs, like orthoferrites~\cite{turov2010symmetry}, (40 nm for $\omega_0 / 2\pi =500~\text{GHz}$, $c=20~\text{km/s}$). The excitation of such short coherent waves is a fundamental problem of modern magnonics since it requires an ultra-compact source of magnons~\cite{hamdi2022terahertz}, despite different finesses, such as the usage of higher-order radial and azimuthal modes.

Here, we propose to employ a spin-current driven domain wall (DW) in an AFM as an ultra-compact source of the propagating coherent SWs. We demonstrate theoretically and by micro-magnetic simulations that the simple spin texture, such as an AFM DW, driven by spin current~\cite{ovcharov2022spin, ovcharov2023antiferromagnetic}, can be a source of the propagating SWs with substantially high frequencies and short wavelengths, comparable to the exchange length of the AFM. We consider a device, schematically shown in Fig. \ref{fig:schema}, which is based on a thin film of an AFM with easy-axis anisotropy and $n$-fold rotational symmetry in the hard plane. In the proposed generator, the spin current flowing from the adjunct layer with the polarization along the principal axis excites the precession of the N\'eel vector within the DW. We assume the finite size of the spin current source with a width $L$ located directly under a DW. The threshold current of the excitation is defined by the value of the anisotropy in the hard plane, and the frequency of the DW precession $\omega$ is tuneable by the strength of the spin current. We show that the above precession of the DW leads to the excitation
of two modes of magnons with the frequencies $(n\pm1)\omega$, where $n$ is the order of the anisotropy. A robust emission of the propagating SWs into the AFM strip occurs when $(n\pm1)\omega>\omega_0$, where $\omega_0$ defines the frequency of AFM resonance. Consequently, the maximum achievable frequency of SWs is $(n+1)\omega$, which corresponds to very short wavelengths of the SW, comparable with the exchange length, especially for the hexagonal AFMs. The excitation of the high wavevectors is possible due to the substantially small width of the DW in AFM, which is hard to achieve by any other method.

\textit{Model.---}The low-energy dynamics of a colinear AFM can be described using the Lagrangian $\mathcal{L} = T - U$ for the N\'eel vector $\mathbf{l}=(\mathbf{M}_1-\mathbf{M}_2)/M_s$, where $|\mathbf{M}_{\text{i}}|=M_s/2$ is the magnetization of the sublattice $i=1,2$ and $M_s$ is the value of saturated AFM magnetization. The ``kinetic'' energy  $T=(M_s/2\gamma \omega_{\text{ex}} ) (\partial_t \mathbf{l})^2$ determines the inertial properties of the AFM spin dynamics, where $\omega_{\text{ex}}= \gamma H_{\text{ex}}$ is the frequency defined by the exchange field $H_{\text{ex}}$ of the AFM, and $\gamma$ is a gyromagnetic ratio.  The ``potential'' term $U=(A/2) (\nabla \mathbf{l})^2 + w_a(\mathbf{l})$ is determined by nonuniform exchange ($A$) and anisotropy energy $w_a$.
Expressing the N\'eel vector in spherical coordinates $\mathbf{l}= \{\sin \theta \cos \phi, \sin \theta \sin \phi, \cos \theta \}$, the anisotropy energy density reads:
\begin{equation}
    w_\text{a} = \frac{K}{2} \sin^2 \theta + \frac{K_\text{n}}{n} \sin^n \theta \sin^2 \left( \frac{n \phi}{2} \right),
    \label{eq:anisotropy}
\end{equation}
where the first term defines uniaxial anisotropy of the easy-axis type ($K>0$), and the second one defines an anisotropy in the hard plane ($K_\text{n}>0$), for an AFM with an n-fold axis $C_{\text{n}}$. Here $z$-axis is chosen along the easy axis of the AFM, $K>K_\text{n}$, and the ground state corresponds to $l_z \rightarrow  \pm 1$, ($\theta = 0, \pi$).

A purely uniaxial AFM model ($K_n = 0$) possesses formal Lorentz invariance~\cite{baryakhtar1983dynamical, galkina2018dynamic} with the characteristic velocity $c=\gamma \sqrt{H_\text{ex} A / M_\text{s}}$ and degeneracy of the antiferromagnetic resonance (AFMR) frequency  $\omega_0=\gamma \sqrt{H_\text{ex} K / M_\text{s}}$. The solution for a stationary DW with boundary conditions $l_z|_{\pm \infty} \rightarrow  \pm 1$ can be found from the minimum of the potential energy $U$ as $\cos \theta_0 = \tanh x / x_0$ and  $\varphi = \varphi_s$, where $x_0=\sqrt{A/K}$ is the thickness of stationary DW and angle $\phi_s$ determines the rotation of the $\mathbf{l}$ vector in the hard plane. The rotational dynamics of interest thus can be described by the transformation $\varphi = \omega t + \varphi_s$ and $x_0 \rightarrow \Delta(\omega)= x_0/ \sqrt{1-\omega^2/\omega_0^2}$, where $\omega$ denotes angular velocity of the N\'eel vector precession in a DW~\cite{baryakhtar1983dynamical, galkina2018dynamic}. To induce the rotational dynamics, a spin current that is polarized along the easy axis of the AFM can be utilized. The frequency $\omega$ dependence on the current $j$ is governed by the equilibrium between the total energy losses in the DW and the energy gained within the constrained region (with the width $L$, see Fig.\ref{fig:schema}) of the spin current's contact area  \cite{ovcharov2022spin}:
\begin{equation}\label{eq:freq_vs_j}
    \alpha \omega = \sigma j \tanh \left( \frac{L}{2 x_0} \sqrt{1 - \frac{\omega^2}{\omega_0^2}} \right),
\end{equation}
where $\alpha$ is an effective Gilbert damping, $\sigma$ is a spin torque efficiency and $j$ is a density of electric current. In the case of a large spin-torque source, $L \gg x_0$ and $\omega \ll \omega_0$, the frequency of the rotation is linearly proportional to the applied current $\omega = \sigma j / \alpha$.

Let us continue our analysis with the second term of Eq. \ref{eq:anisotropy} -- namely, anisotropy in the hard plane $K_\text{n}$, that leads to the excitation of spin waves.
In order to investigate spin waves excitation, we consider small perturbations of the initial DW solution as $\theta = \theta_0(x) + \vartheta(x, t)$ and $\varphi = \varphi_s + \omega t + \mu(x, t) / \sin \theta_0 (x)$. It is convenient to combine polar and azimuthal perturbations into a single complex variable $\psi = \mu  + i \vartheta$. Assuming small value of the symmetry-reducing term $K_n \ll K$, one can obtain the linearized equation for $\psi$ in the form~\cite{galkina2017precessional} (see Supplementary Material \cite{[{See Supplemental Material at }]supplemental} for the details):
\begin{equation}\label{eq:main}
\begin{split}
    \hat{H}_0 \psi + \frac{1}{\omega_0^2-\omega^2} \partial^2_t \psi - \frac{2 i \omega \cos \theta_0}{\omega_0^2-\omega^2} \partial_t \psi = \\
     B_n^+ (\xi) e^{i n \omega t} + B_n^- (\xi) e^{-i n \omega t},
\end{split}
\end{equation}
where $\hat{H}_0$ is the Schr\"odinger operator with the reflectionless P\"oschl-Teller potential $\hat{H}_0 = - \partial^2_{\xi} + 1 - 2 \cosh^2 \xi$, $\xi = x/\Delta$. The left side of Eq. (\ref{eq:main}) describes small-amplitude excitations in an AFM containing a precessing DW. The included type of potential created by a DW for linear SWs was already discussed for the FM as well as AFM materials. Particularly, the emission of exchange SWs from a Bloch DW, excited by a microwave magnetic field, was predicted for FMs in \cite{whitehead2017theory}. For uniaxial AFMs, the above approach was employed in \cite{kim2014propulsion}, where the propulsion of a DW by incoming SWs was demonstrated. The right-hand side of Eq. (\ref{eq:main}) represents periodic driving ``force" with frequencies $\pm n\omega t$ and corresponding amplitudes $B_n^{\pm} (\xi) = - i B_n \sin^{n-1} \theta_0 ( \cos \theta_0 \pm \sin \theta_0)/2$, where $B_n = K_n / 2 K (1-\omega^2/\omega_0^2)$ is proportional to the value of the anisotropy in the hard plane. Please note that superscripts indicate the sign of the corresponding frequency, i.e., the direction of $\psi$ rotation.

The free solution of Eq. (\ref{eq:main}) can be represented in the form of a planar wave $ \psi = e^{i(\Tilde{k} \xi + \Tilde{\Omega} t)}$, where $\Tilde{k}=k \Delta$ is the rescaled wave vector and $ \Tilde{\Omega} = \Omega \pm \omega$ is the wave frequency in the rotating reference frame. At a large distance from the DW, $\Tilde{k}$ and $\Tilde{\Omega}$ are connected by the relation
\begin{equation} \label{eq:kw_correlation}
    \Tilde{k}^2|_{\xi \rightarrow \pm \infty} =  \frac{(\omega \mp \Tilde{\Omega})^2 - \omega^2_0 }{ \omega^2_0 - \omega^2},
\end{equation}
which is a transformed version of the known dispersion law for spin waves $\Omega^2 = \omega_0^2 + c^2 k^2$ in the observer's coordinate system.

As it follows from the right-hand side of the Eq. (\ref{eq:main}), $\psi(\xi, t)$ should be expressed as a linear combination of terms with both positive and negative frequencies $ \pm n\omega t$. However, it is sufficient to consider one frequency sign since the part with the opposite sign is symmetric with respect to $\xi=0$. Separating spatial and time variables as  $\psi=\chi^{\pm}(\xi)e^{\pm i n \omega t}$ the equation (\ref{eq:main}) can be written as

\begin{equation}\label{eq:chi}
    - \partial^2_{\xi} \chi^{\pm} + U^{\pm}_{\omega}(\xi)\chi^{\pm} = B_n^{\pm} (\xi),
\end{equation}
for the spatial part $\chi^{\pm}(\xi)$, where $U^{\pm}_{\omega}(\xi)$ is a dimensionless potential for SWs created by a DW rotation and is given by
\begin{equation}\label{eq:potential}
    U^{\pm}_{\omega}(\xi) = 1 - \frac{2}{\cosh^2 \xi} - \frac{n^2\omega^2}{\omega_0^2 - \omega^2} \pm \frac{2 n \omega^2}{\omega_0^2-\omega^2} \tanh \xi
\end{equation}

The function $B_n^{\pm} (\xi)$ in Eq.~\ref{eq:chi} defines the amplitude of a spin wave, while the potential $U^{\pm}_{\omega}(\xi)$ defines the condition for its propagation. Particularly, for the propagating SW in the form $\chi(\xi)\propto e^{\pm i \Tilde{k}\xi}$ the wavevector acquires the real value $\Tilde{k}^2 > 0$ when
\begin{equation}\label{eq:k_cond}
    U^{\pm}_{\omega}(\pm \infty) < 0.
\end{equation}
Thus, Eq. (\ref{eq:k_cond}) is a condition for a SW propagation with a wavevector $\Tilde{k}^2$ given by the relation (\ref{eq:kw_correlation}) with a substitution $\Tilde{\Omega} \rightarrow n\omega$. 

\begin{figure}[hbt!]
\centering
	\includegraphics[width=\linewidth]{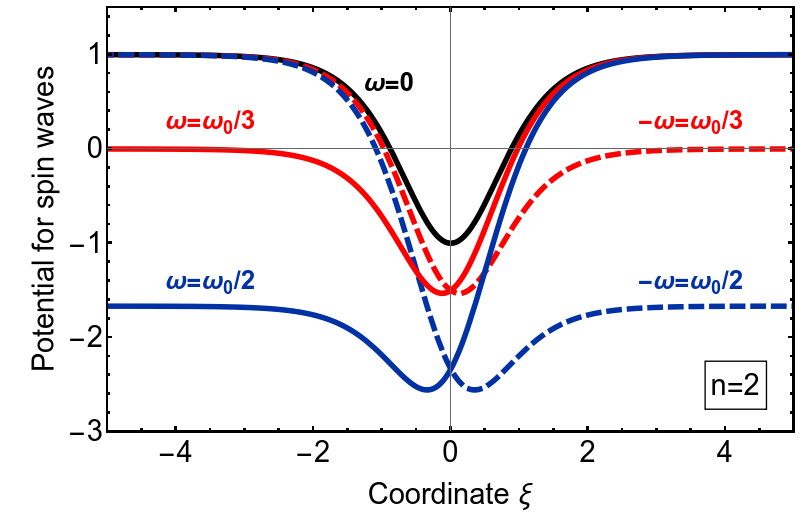}
	\caption{Potential $U^{\pm}_{\omega}$ created by a precessing DW for SWs, given by Eq. (\ref{eq:potential}), for a different frequency $\omega$ of a DW precession. Solid lines correspond to the positive sign of a SW frequency, while dashed lines correspond to the negative sign.}
	\label{fig:potential}
\end{figure}

The potential $U^{\pm}_{\omega}$ depends on the frequency of DW rotation $\omega$, see Fig. \ref{fig:potential}, which in turn can be controlled by the applied current in accordance with Eq. \ref{eq:freq_vs_j}. Thus, by increasing the current, the condition for the emission is fulfilled when certain critical frequencies are exceeded:
\begin{equation}\label{eq:w_cond}
   \omega>\omega_\text{cr}, \quad \omega_\text{cr}^2=\omega_0^2/(n \pm 1)^2.
\end{equation}

In general, critical frequencies (\ref{eq:w_cond}) distinguish three frequency ranges of the DW precession. At low frequencies $\omega<\omega_0 / (n + 1)$ propagating SWs are not excited, since the wavevector $k$ is purely imaginary. At $\omega_0 / (n + 1)<\omega<\omega_0 / (n - 1)$ only one branch of propagating SWs is emitted by a DW with a frequency $\Omega=(n+1)\omega$. At $\omega>\omega_0 / (n - 1)$ the second branch of SWs appears with frequency $\Omega=(n - 1)\omega$. The third region is, however, absent for two-fold anisotropy with $n=2$, since corresponding frequencies lay below AFMR. 

\textit{Micromagnetic simulations.---}To validate our analytical findings, we carried out micro-magnetic simulations using \emph{MuMax3} solver~\cite{vansteenkiste2014design} for a system schematically shown in Fig.~\ref{fig:schema}. 
The AFM film has lateral sizes $0.14 \times 1.56$~$\mu$m$^2$ with a thickness of 5~nm. The selected
parameters used for the AFM correspond to the DyFeO$_3$ and are given as~\cite{turov2010symmetry}: $\alpha=10^{-3}$, $M_s = 8.4 \cdot 10^5$~A/m, $A=18.9$~pJ/m, $H_{ex}=670$~T, the anisotropy constant along the easy axis $K = 300$~kJ/m$^3$. These parameters correspond to the characteristic speed $c=22$~km/s, the frequency of the AFM resonance $\omega_0/2\pi = 0.45$~THz, and the width of a stationary DW $x_0=8$~nm. DyFeO$_3$ was chosen due to the relatively simple tunability of the second anisotropy $K_2$ in this material, for example, by temperature \cite{koshizuka1988raman, wijn19945, fu2014terahertz}. Particularly, at low temperatures, it is possible to achieve a uniaxial state \cite{koshizuka1988raman} with $K_n=0$, where two magnon modes are degenerated, and by varying temperature in the vicinity of this point, it is possible to tune $K_2$ in a wide range.

A spin current source with a lateral size $0.14 \times 0.1$~$\mu$m$^2$ ($L=100$~nm) is positioned under a DW at the center of the device, injecting a spin torque polarized along the easy axis of the AFM. The frequency of the DW rotation is evaluated at the central location, where an initially relaxed DW is present. The selected polarization of a spin torque does not induce translational DW movement. However, additional methods, such as nanoconstriction-based pinning, can be utilized for the DW stabilization, if necessary. The frequencies of the excited SWs are measured at a distance of $300$~nm from the DW.

\begin{figure}[hbt!]
\centering
	\includegraphics[width=0.8\linewidth]{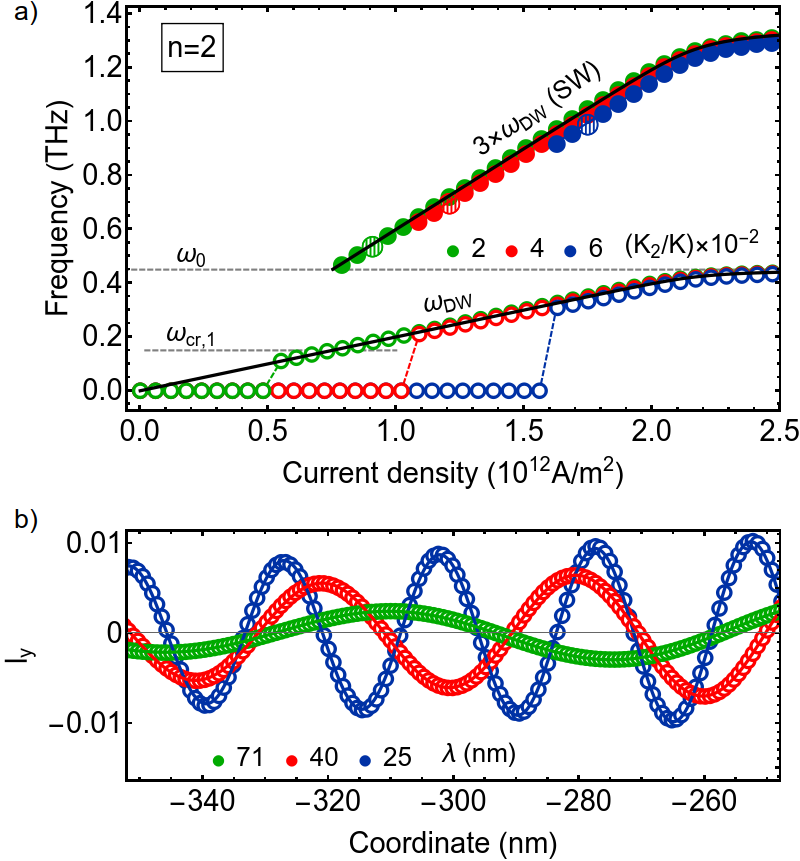}
	\caption{The results of simulations for two-fold anisotropy in a hard AFM plane, $n=2$. a) The frequency of the DW rotation (empty circles) and emitted SWs (filled circles) are shown as a function of the applied current density for different values of $K_2$ anisotropy. Solid black lines are calculated analytically using Eq. (\ref{eq:freq_vs_j}). Angular frequency labels are employed for simplicity and correspond to the respective rotational frequencies $f=\omega/2\pi$ b) The profiles of emitted SWs far from a DW with the extracted values of the wavelengths: $\lambda=71$ nm (red), 40 nm (green), 25 nm (blue).}
	\label{fig:n2}
\end{figure}

Figure~\ref{fig:n2} shows the results of simulations with 2-fold anisotropy in the hard AFM plane, considering various values of $K_2$. The presence of $K_2$ anisotropy induces the excitation threshold current $\sigma j_\text{th}=\omega_2^2/(2\omega_{ex})$, where $\omega_n=\gamma \sqrt{H_\text{ex} K_n / M_s }$. In particular, for $K_2/K = 0.02$, the excitation starts at $j_\text{th}=0.53\times10^{12}$~A/m$^2$ with frequency $\omega_\text{th} \simeq \sigma j_\text{th}/\alpha = \omega_n^2/(2 \alpha \omega_{ex}) \approx 100$~GHz. With an increase in current, the frequency of DW rotation reaches the critical frequency $\omega_\text{cr,1}=\omega_0/3=150$~GHz, leading to the detection of SWs at a large distance from the DW. The dependence of the DW frequency on the applied current is in good agreement with Eq. \ref{eq:freq_vs_j}, despite that it is derived with the assumption $K_n/K \ll 1$. As predicted above, the frequency of the propagating SW is multiple of the DW frequency with a factor of $n+1=3$ and, hence, follows the scaled dependence (\ref{eq:freq_vs_j}) on the applied current.

Since increasing the anisotropy $K_n$ leads to an increase in the threshold current $j_{th}$, the frequency of a DW precession at the threshold exceeds $\omega_\text{cr,1}$ for high values of $K_n$. As a result, only SWs with a substantial frequency gap above AFMR can be excited in this case, see $K_n/K=0.04$ and $0.06$ in Fig.~\ref{fig:n2}. Another outcome of raising $K_n$ is the increase of the SW amplitude, since driving term $ B_n^{\pm}\propto K_n$ in Eq. (\ref{eq:chi}). The SW radiation serves as an additional dissipation mechanism and results in a reduction of the measured DW frequency (and correspondingly the frequency of emitted SWs) as compared to the dependency (\ref{eq:freq_vs_j}). This effect is visible in Fig.~\ref{fig:n2} for $K_n/K=0.06$.

\begin{figure}[t!]
\centering
	\includegraphics[width=0.8\linewidth]{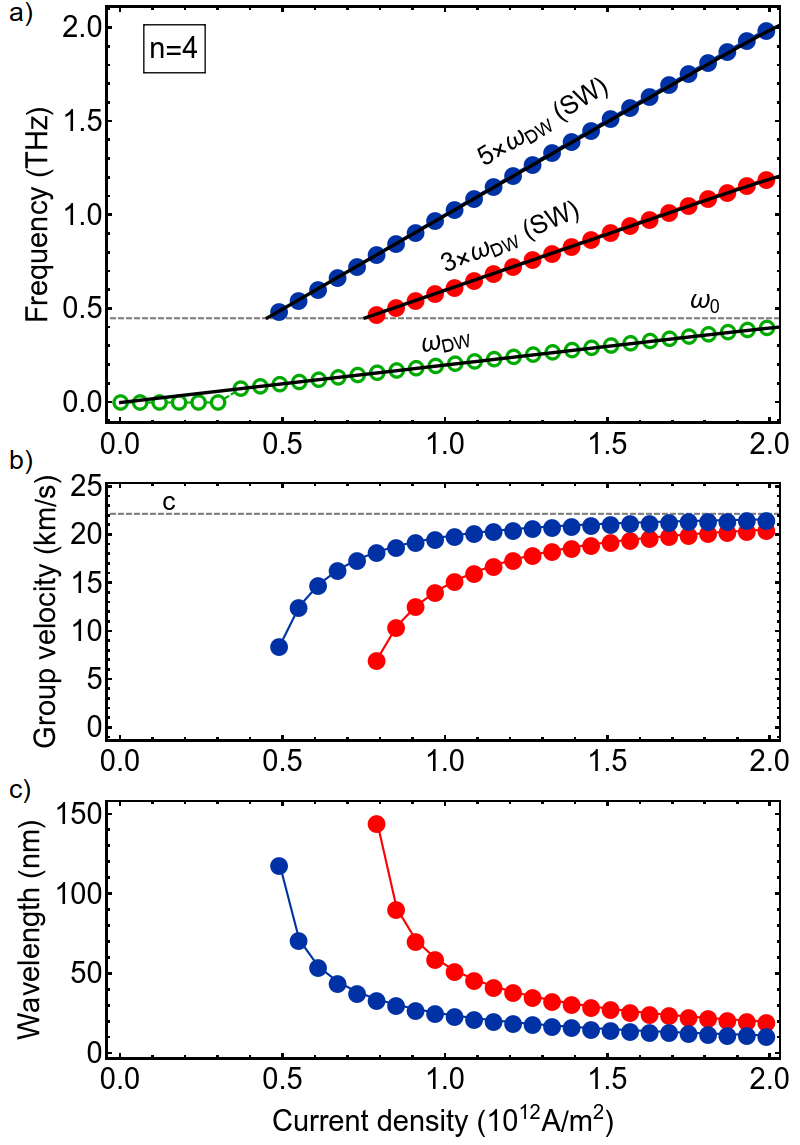}
	\caption{The results of simulations for four-fold anisotropy in a hard AFM plane, $n=4$. a) The frequency of the DW rotation (empty circles) and emitted SWs (filled circles) are shown as a function of the applied current density for $K_4/K=2\times10^{-2}$. Solid black lines are calculated analytically using Eq. (\ref{eq:freq_vs_j}). b) Group velocity and c) wavelength of the excited SWs as a function of the applied current density. }
	\label{fig:n4}
\end{figure}

The results for 4-fold anisotropy with $K_4/K=0.02$ are shown in Fig.~\ref{fig:n4}. Here, all other parameters are left unchanged for the possibility of a direct comparison with the $n=2$ case. The excitation threshold current for $n=4$ is given by $\sigma j_\text{th} = \omega_4^2/(3\omega_\text{ex})$, which corresponds to $j_\text{th}=0.35\times 10^{12}$ A/m$^2$ and a frequency of $\omega_{th}/2\pi=70$ GHz. Upon reaching the first critical frequency $\omega_\text{cr,1}=\omega_0/5=90$ GHz, only SWs with a five-fold frequency are observed. As the current is further increased, the DW surpasses the subsequent critical frequency $\omega_\text{cr,2}=\omega_0/3=150$ GHz, leading to the emission of SWs with a triple frequency as well.

Fig. \ref{fig:n4} b) and c) show group velocity and a wavelength of emitted SWs as a function of applied current. Our simulation results suggest ultra-high velocities, exceeding 10km/s, even at low supercriticality, while at higher currents, the velocity of emitted SWs is closely approaching the maximum value of 22km/s. Such a result is extremely hard to achieve by any other method of excitation due to the extremely small wavelength $\simeq 10$nm (see Fig. \ref{fig:n4} c) of the magnon.

\textit{Discussion.---} One can note that Eq. (\ref{eq:main}) is derived for the conservative case, i.e., does not take dissipation and spin current into account. Thus, the emission of the SWs can be created by any mechanism, which leads to the corresponding spin precession in the DW, and spin torque induced by a current is one of them. Gilbert damping defines the frequency of the DW precession in accordance to Eq. (\ref{eq:freq_vs_j}) and also leads to the decay of the propagating SWs, as one can see in the inset of Fig. \ref{fig:n2}.

It's worth mentioning that the anisotropy in the hard plane is not the only mechanism that leads to the reduction of the DW dynamic symmetry~\cite{galkina2017precessional}. The corresponding effect of the SWs emission can occur in AFM with a specific form of the Dzyaloshinskii–Moriya interaction (DMI) characterized by a function $D(\theta, \phi)$. The forms of the functions $D(\theta, \phi)$ for many AFMs are detailed in Ref.~\cite{gomonai1990symmetry}. The incorporation of DMI results leads to the term of a form $D(\theta) \sin n \omega t$ in the right-hand side of Eq.~(\ref{eq:main}), which acts as a periodic driving ``force'', similarly to the effect of anisotropy.

To summarize, it has been shown theoretically and confirmed by micromagnetic simulations that the AFM DW, in which internal rotational dynamics is excited by a spin current, can be utilized as a generator of SWs with remarkably high frequencies and group velocities, which correspond to short wavelengths of the order of the AFM exchange length. The AFM DW, due to its small characteristic width, serves as an efficient generator of SWs that are difficult to excite by other methods. In addition, the application of such radiation for the synchronization of AFM oscillators with multiple DWs, where the dynamics is induced by the spin torque, is of particular interest.

\hfill

This project is partly funded by the European Research Council (ERC) under the European Union’s Horizon
2020 research and innovation programme (Grant TOPSPIN No 835068) and the Swedish Research Council Framework Grant Dnr. 2016-05980.

\bibliography{main}

\begin{thebibliography}{43}
\bibitem [{sup()}]{supplemental}%
  \BibitemOpen
  \href@noop {} {}\bibinfo {howpublished} {[URL will be inserted by a publisher]}\BibitemShut {NoStop}%
\end{thebibliography}

\end{document}



\title[]{Supplementary material\\ for \\Emission of fast-propagating spin waves by \\an antiferromagnetic domain wall driven by spin current}

\author{Roman V. Ovcharov}
\affiliation{ 
Department of Physics, University of Gothenburg, Gothenburg 41296, Sweden
}

\author{B.A. Ivanov}
\affiliation{
Institute of Magnetism of NASU and MESU, Kyiv 03142, Ukraine
}
\affiliation{
William H. Miller III Department of Physics and Astronomy, Johns Hopkins University, Baltimore, Maryland 21218, USA
}

\author{Johan \AA kerman}
\affiliation{ 
Department of Physics, University of Gothenburg, Gothenburg 41296, Sweden
}
\affiliation{
Center for Science and Innovation in Spintronics, Tohoku University, Sendai 980-8577, Japan
}
\affiliation{
Research Institute of Electrical Communication, Tohoku University, Sendai 980-8577, Japan
}

\author{Roman S.  Khymyn}%
\affiliation{ 
Department of Physics, University of Gothenburg, Gothenburg 41296, Sweden
}

\maketitle

The Lagrangian density of the sigma-model describing the dynamics of the N\'eel vector $\mathbf{l}$ in the nondissipative limit can be represented in the form:
\begin{equation}
\mathcal{L} =  \frac{M_s}{2\gamma \omega_{ex}} (\partial_t \mathbf{l})^2  - \frac{A}{2}(\nabla \mathbf{l})^2 - w_a(\mathbf{l}),
\end{equation}
where $M_s$ is the saturation magnetizaton of the AFM,  $\gamma$ is a gyromagnetic ratio, $\omega_{ex}=\gamma H_{ex}$ is the frequency defined by the uniform exchange field $H_{ex}$, $A$ is the inhomogeneous exchange constant. The anisotropy energy in spherical coordinates  $\mathbf{l}=\{ \sin \theta \cos \phi, \sin \theta \sin \phi, \cos \theta \}$ is taken in the form
\begin{equation}
w_a(l) = \frac{K}{2}\sin^2 \theta + \frac{K_n}{n}\sin^n \theta \sin^2 \left( \frac{n\phi}{2} \right).
\end{equation}
Here, the first term defines purely uniaxial anisotropy of the easy-axis type $K>0$, $z$ axis is chosen along the easy axis of the AFM so that the ground state corresponds to $\theta = 0, \pi$. The second term defines the simplest form of anisotropy in the basal plane for an AFM with an easy axis of the $n$-th order $C_n$, which is assumed to be small $K_n \ll K$.

For the Lorentz-invariant model of an AFM in the approximation of purely uniaxial symmetry $K_n = 0$, the solution for a stationary wall has the form 
\begin{equation}
\cos \theta_0 = \tanh \left( \frac{x}{x_0} \right), \quad \phi = \phi_0,
\end{equation}
where the  angle $\phi_0$ determines  the  rotation  plane  of  the N\'eel vector $\mathbf{l}$ in the wall. The characteristic scale is given by the thickness of the DW at rest $x_0=\sqrt{A/K}$, which can be expressed through the characteristic speed of magnons $c=\gamma \sqrt{H_{ex}A/M_s}$ and characteristic frequency (the magnon gap defined by the easy axis anisotropy) $\omega_0=\gamma \sqrt{H_{ex}K/M_s}$ as $x_0 = c / \omega_0$. The characteristic intrawall frequency is defined by the anisotropy in the basal plane $\omega_n = \gamma \sqrt{H_\text{ex} K_n / M_\text{s}}$. The DW with the precession of spins inside is described by
\begin{equation}\label{eq:rotating_profile}
   \cos \theta_0 = \tanh \left( \frac{x}{\Delta} \right), \quad \phi = \phi_0 + \omega t, 
\end{equation}
where  the DW thickness $\Delta$ has the next ansatz $\Delta = x_0/\sqrt{1-\dot{\phi}^2/\omega_0^2}= x_0/\sqrt{1-\omega^2/\omega_0^2}$.

The dissipative function, taking into account the contribution of spin torque (ST), is given by:
\begin{equation}
    \mathcal{R} = \frac{M_s}{2 \gamma} \left[ \alpha (\partial_t \mathbf{l})^2 - 2 \tau \partial_t \mathbf{l}  (\mathbf{l} \times \mathbf{p}) \right],
\end{equation}
where $\alpha$ is the effective dissipation constant and $\tau=\sigma j$ is the torque created by a spin current expressed in the units of frequency, $\sigma$ is the torque-current proportionality coefficient. 

\section{Collective coordinates approach.}

The Lagrangian density and the rate of energy change associated with the rotational dynamics are given  in spherical coordinates as

\begin{equation}\label{eq:lagrangian_spherical}
    \mathcal{L}_{\phi} = \frac{M_s}{2\gamma \omega_{ex}} \left[ (\dot{\phi}^2 -\omega_0^2) \sin^2 \theta - \frac{2 \omega_n^2}{n} \sin^n \theta \sin \left( \frac{n \phi}{2} \right) \right],
\end{equation}

\begin{equation}\label{eq:dote_spherical}
    \dot{E}_{\phi} =  - \dot{\mathbf{l}} \frac{\partial \mathcal{R}}{ \partial \dot{\mathbf{l}} } = - \frac{\alpha M_s}{\gamma} \dot{\phi}^2 \sin^2 \theta + \frac{\tau M_s}{\gamma} \dot{\phi} \sin^2 \theta.
\end{equation}

The forced rotational dynamics of an AFM DW can be described by the collective coordinate $\Phi(t)$. The effective Lagrangian density and energy dissipation function can be obtained by integrating (\ref{eq:lagrangian_spherical}) and (\ref{eq:dote_spherical}) over $x$ in infinite limits using the DW profile (\ref{eq:rotating_profile}) with a substitution $\phi \rightarrow \Phi$. Energy dissipation function can be written as $\dot{E}_\Phi = \dot{\Phi} F_\Phi$, and we are interested in the nonconservative moment of force $F_\Phi$ that will be incorporated into a standard Euler-Lagrange equation.
\begin{equation}
    F_\Phi = - \frac{2 \alpha M_s\Delta}{\gamma} \dot{\Phi} + \frac{2 \tau M_s \Delta}{\gamma}.
\end{equation}

The ``kinetic'' part $T_\Phi$ of the Lagrangian density $\mathcal{L}_\Phi = T_\Phi - U_\Phi$ is equal to $T_\Phi=(M_s\Delta/\gamma \omega_{ex})\dot{\Phi}^2$, and the potential part depends on the symmetry order $n$:

\begin{equation*}
\frac{ \gamma \omega_{ex}}{ M_s \Delta } U_\Phi= 
\begin{cases}
\omega_2^2 \sin^2 \Phi & \text{if $n=2$}\\
(\omega_4^2 / 3) \sin^2 2 \Phi & \text{if $n=4$}
\end{cases}
\end{equation*}

Therefore, for $n=2$, the equation of motion reads
\begin{equation}
    \frac{ 1 }{  \omega_{ex}}  \ddot{\Phi} + \alpha \dot{\Phi}  +  \frac{ \omega_2^2}{ 2 \omega_{ex}}   \sin 2 \Phi =  \tau,
\end{equation}
that specifies the threshold torque as $\tau_{th} = \sigma j_{th} =  \omega_2^2 / (2 \omega_{ex})$. For $n=4$, one can obtain
\begin{equation}
    \frac{ 1}{  \omega_{ex}} \ddot{\Phi} + \alpha \dot{\Phi}  +   \frac{ \omega_4^2}{3  \omega_{ex}}   \sin 4 \Phi=   \tau,
\end{equation}
with the threshold torque $\tau_{th}=  \omega_4^2 / (3 \omega_{ex})$.

\section{Perturbation theory.}

The complete Lagrangian in a form convenient for analyzing the one-dimensional DW with $\theta = \theta(x,t)$ and $\phi = \phi(x,t)$ reads
\begin{equation}
\begin{split}
L = & E_0 \int \frac{dx}{2 x_0} \left\{ \omega_0^{-2} \left[ \left( \partial_t \theta\right)^2 + \left( \partial_t \phi\right)^2 \sin^2\theta  \right]   
 - x_0^2 \left[ \left( \partial_x \theta \right)^2 +  \left(\partial_x \phi \right)^2 \sin^2\theta \right] - \sin^2 \theta -  \frac{2 K_n}{nK}\sin^n \theta \sin^2 \left( \frac{n\phi}{2}\right) \right\},
\end{split}
\end{equation}
where $E_0=S_{\perp}\sqrt{AK}$ is the energy of a planar stationary wall with the cross-sectional area of the AFM sample $S_{\perp}$. Equations  for  the  variables $\theta$ and $\phi$ can  be  written  as follow:
\begin{equation}\label{eq:theta}
\omega_0^{-2} \partial^2_{t} \theta  - x_0^2 \partial^2_{x} \theta  + \sin \theta \cos \theta  \left[ 1 + x_0^2 ( \partial_x \phi)^2 - \omega_0^{-2}  (\partial_t \phi)^2  \right] = \frac{K_n}{K} \sin^{n-1}\theta \cos \theta \sin \left(\frac{n \phi}{2}\right),
\end{equation}
\begin{equation}\label{eq:phi}
\omega_0^{-2} \partial_t \left( \sin^2 \theta \partial_t \phi \right) - x_0^2 \partial_x \left( \sin^2 \theta \partial_x \phi \right) = \frac{K_n}{2 K} \sin^n \theta \sin(n\phi).
\end{equation}

The system (\ref{eq:theta}) and (\ref{eq:phi}) without anisotropy in the basal plane has the solution of the DW with the precession of spins~(\ref{eq:rotating_profile}). Considering that the terms on the right-hand side are small $K_n \ll K$, we will consider  small  perturbations  to  the  ``zero'' solution (\ref{eq:rotating_profile}) in the form:
\begin{equation}
\begin{split}
& \theta = \theta_0(x) + \vartheta(x, t),\\
& \phi = \phi_0 + \omega t + \mu(x, t)/ \sin \theta_0(x).
\end{split}
\end{equation}

Linearizing the left-hand sides of the system (\ref{eq:theta}) and (\ref{eq:phi}) with respect to $\vartheta$ and $\mu$ and limiting to the zeroth approximation in the right-hand sides, one obtains a system of linear inhomogeneous equations for $\vartheta$ and $\mu$ in the following form:
\begin{equation}
\begin{aligned}
& \left( \hat{H}_0 + \frac{1}{\omega^2_0 - \omega^2} \partial^2_t \right) \vartheta  - \frac{2\omega \cos \theta_0}{\omega^2_0 - \omega^2}\partial_t \mu = - B_n\sin^{n-1} \theta_0 \cos \theta_0 \cos n \omega t,\\
& \left( \hat{H}_0 + \frac{1}{\omega^2_0 - \omega^2} \partial^2_t \right) \mu  + \frac{2\omega \cos \theta_0}{\omega^2_0 - \omega^2} \partial_t \vartheta = B_n \sin^n \theta_0 \sin n \omega t,
\end{aligned}  
\end{equation}
Here, the terms that do not contain time dependencies are omitted, and the following notation is used $B_n = K_n / 2K ( 1 - \omega^2/ \omega^2_0)$. The equations for $\vartheta$ and $\mu$ contain the Schr\"odinger operator with the P\"oschl-Teller potential
\begin{equation}
\hat{H}_0 = - \partial^2_\xi + 1 - \frac{2}{\cosh^2 \xi}, \quad \xi = x / \Delta
\end{equation}

For further analysis, it is convenient to introduce a complex variable  $\psi = \mu + i \vartheta$. By using the transformation
\begin{equation}
    \begin{aligned}
     B_n \sin^{n-1} \theta_0 & ( \sin \theta_0 \sin n \omega t - i \cos \theta_0 \cos n \omega t )= \\
    &- \frac{i B_n \sin^{n-1}\theta_0}{2} \left[ (\cos \theta_0 +  \sin \theta_0  ) e^{i n \omega t} + (\cos \theta_0 - \sin \theta_0)  e^{- i n \omega t}  \right] = \\ 
    &- \frac{i B_n \sech^{n-1}\xi}{2} \left[ (\tanh \xi +  \sech \xi  ) e^{i n \omega t} + (\tanh \xi -  \sech \xi )  e^{- i n \omega t}  \right] = B_n^+ (\xi) e^{i n \omega t} + B_n^- (\xi) e^{-i n \omega t},
    \end{aligned}
\end{equation}
one gets the equation for $\psi$ in the following form:
\begin{equation}\label{sup:eq:main}
    \hat{H}_0 \psi + \frac{1}{\omega_0^2-\omega^2} \partial^2_t \psi - \frac{2 i \omega \cos \theta_0}{\omega_0^2-\omega^2} \partial_t \psi = B_n^+ (\xi) e^{i n \omega t} + B_n^- (\xi) e^{-i n \omega t}.
\end{equation}

For a free solution of the form $\psi \propto e^{i( \Tilde{k} \xi + \Tilde{\Omega} t )}$,  the Lorentz-transformed wave vector $\Tilde{k}=k\Delta = k x_0 / \sqrt{1 - \omega^2 / \omega_0^2}$ and the wave frequency in the rotating frame $\Tilde{\Omega}$ are connected at a distance from the DW by the reletation
\begin{equation}\label{sup:eq:komega}
    \Tilde{k}^2|_{\xi \rightarrow \pm \infty} = \frac{( \omega \mp \Tilde{\Omega})^2 - \omega_0^2}{\omega_0^2 - \omega^2}
\end{equation}

From Eq. (\ref{sup:eq:main}), the solution for $\psi$ should include additives with positive and negative frequencies $\pm n \omega t$. The general solution can be represented as a superposition of solutions with different frequency signs. It is sufficient to consider the part of the equation that is proportional to $ e^{i n \omega t} $, since the solution that is proportional to $ e^{- i n \omega t} $ is symmetric with respect to $ \xi = 0 $. The equation for the coordinate part of the function $\psi = \chi(\xi) e^{+ i n \omega t}$ reads
\begin{equation}
    \begin{aligned}
    -\partial^2_\xi \chi + \left( 1 -\frac{2}{\cosh^2 \xi} - \frac{n^2 \omega^2}{\omega^2_0 - \omega^2} + \frac{2 n \omega^2}{\omega^2_0 - \omega^2} \tanh \xi \right) \chi =  B_n^+ (\xi)
    \end{aligned}
\end{equation}

To simplify, we replace the expressions of type $ \sin^n \theta_0 $ by the delta function $ \sin ^ n \theta_0 \xrightarrow{} u_n \delta (\xi) $, and the expressions of type $ \sin ^ {n-1} \theta_0 \cos \theta_0 $ by the derivative of the delta function with respect to the coordinate $ \sin ^ {n-1} \theta_0 \cos \theta_0 \xrightarrow{} u_ {n-1} \delta ^ {'} (\xi) / (n-1) $ , where the constant $ u_n $ is determined from the condition $ \int _ {- \infty} ^ {\infty} \sin ^ n \theta_0 dx = \int _ {- \infty} ^ {\infty} \sech^n \xi d\xi = u_n $, and we also approximate $ \cos \theta_0 $ by a step function $\Xi$, putting $ \Xi = \sign \xi$. Formally, these simplifications are only applicable for very large $ k >> 1 $; however, the numerical calculation showed that the results obtained in this approximation are qualitatively applicable to $ k \sim 1 $.

\begin{equation}
    -\partial^2_\xi \chi + \frac{\omega_0^2 - \omega^2(1 -  n \Xi)^2}{\omega_0^2 - \omega^2} \chi = - \frac{i B_n}{2} \left[ u_n \delta(\xi) + \frac{u_{n-1}}{n-1} \delta^{'}(\xi) \right]
\end{equation}

The solution of the coordinate part thus can be found as a superposition of the solutions with separately considered delta function and a derivative of the delta function in the form:
\begin{equation}
    \chi(\xi) = 
    \begin{aligned}
      \begin{cases} 
         L e^{- i k_{(-)} \xi}  & \text{at $\xi < 0$},\\
         R e^{  i k_{(+)} \xi}  & \text{at $\xi > 0$},
     \end{cases}
     \end{aligned}
\end{equation}
where the quantities $ k_{(+)} $ and $ k_{(-)} $ are given by substituting $\Tilde{\Omega} \rightarrow n \omega$ into Eq.~(\ref{sup:eq:komega}):
\begin{equation}\label{eq:kw}
    k^2_{(\pm)} = \Tilde{k}^2|_{\xi \rightarrow \pm \infty} = \frac{(n \mp 1)^2\omega^2 - \omega_0^2 }{\omega_0^2 - \omega^2} 
\end{equation}

The contributions to the amplitude can be found from the join conditions:
\begin{equation}
\begin{cases} 
     \chi(-0) - \chi(+0) = - \frac{i B_n u_{n-1}}{2 (n-1)} ,\\
     \chi^{'}(-0) - \chi^{'}(+0) = - \frac{i B_n u_n}{2} 
 \end{cases} 
 \end{equation}